\RequirePackage{fix-cm}
\documentclass[natbib,twocolumn]{svjour3}          

\usepackage{graphicx}
\usepackage{flushend}

\usepackage{marvosym}
\newcommand{\envelope}{(\hspace{1pt}\raisebox{-.5pt}{\scalebox{1.45}{\Letter}}\hspace{1pt})}

 \usepackage{mathptmx}      
\usepackage{textcomp}

\usepackage{etoolbox}
\makeatletter
\patchcmd{\NAT@citex}
  {\@citea\NAT@hyper@{%
     \NAT@nmfmt{\NAT@nm}%
     \hyper@natlinkbreak{\NAT@aysep\NAT@spacechar}{\@citeb\@extra@b@citeb}%
     \NAT@date}}
  {\@citea\NAT@nmfmt{\NAT@nm}%
   \NAT@aysep\NAT@spacechar\NAT@hyper@{\NAT@date}}{}{}

\patchcmd{\NAT@citex}
  {\@citea\NAT@hyper@{%
     \NAT@nmfmt{\NAT@nm}%
     \hyper@natlinkbreak{\NAT@spacechar\NAT@@open\if*#1*\else#1\NAT@spacechar\fi}%
       {\@citeb\@extra@b@citeb}%
     \NAT@date}}
  {\@citea\NAT@nmfmt{\NAT@nm}%
   \NAT@spacechar\NAT@@open\if*#1*\else#1\NAT@spacechar\fi\NAT@hyper@{\NAT@date}}
  {}{}
\makeatother

\usepackage[colorlinks, citecolor=blue, linkcolor=blue, urlcolor  = blue]{hyperref}

 \journalname{Rheologica Acta}

\begin{document}
\sloppy

\title{Restructuring and aging in a capillary suspension
}


\author{Erin Koos \and Wolfgang Kannowade \and Norbert Willenbacher 
}


\institute{E. Koos \envelope \and Wolfgang Kannowade \and Norbert Willenbacher 
	\at Institute for Mechanical Process Engineering and Mechanics, Karlsruhe Institute of Technology (KIT), Stra{\ss}e am Forum 8, 76131 Karlsruhe, Germany \\
	\email{erin.koos@kit.edu}             \\
             \emph{Present address of W. Kannowade:} Admedes Schuessler, L\"{o}wentorstra{\ss}e 30, 70179 Stuttgart, Germany.  
}

\date{Received: 21 May 2014 / Revised: 30 September 2014 / Accepted: 30 September 2014\\
{\textcopyright} Springer-Verlag Berlin Heidelberg 2014}

\maketitle

\begin{abstract}
The rheological properties of capillary suspensions, suspensions with small amounts of an added immiscible fluid, are dramatically altered with the addition of the secondary fluid. We investigate a capillary suspension to determine how the network ages and restructures at rest and under applied external shear deformation. The present work uses calcium carbonate suspended in silicone oil (11~\% solids) with added water as a model system. Aging of capillary suspensions and their response to applied oscillatory shear is distinctly different from particulate gels dominated by the van der Waals forces.  The suspensions dominated by the capillary force are very sensitive to oscillatory flow, with the linear viscoelastic regime ending at a deformation of only 0.1~\% and demonstrating power-law aging behavior.  This aging persists for long times at low deformations or for shorter times with a sudden decrease in the strength at higher deformations. This aging behavior suggests that the network is able to rearrange and even rupture. This same sensitivity is not demonstrated in shear flow where very high shear rates are required to rupture the agglomerates returning the apparent viscosity of capillary suspensions to the same viscosity as for the pure vdW suspension. A transitional region is also present at intermediate water contents wherein the material response depends very strongly on the type, strength, and duration of the external forcing. 
\end{abstract}

\section*{Introduction} \label{intro}

Capillary suspensions are formed by adding a small amount of a secondary liquid to the continuous phase of a suspension \citep{Koos:2011}.  The addition of an immiscible fluid either reinforces an existing space-spanning network created by, e.g., the van der Waals force or can even create a network when the particles are well stabilized \citep{Kao:1975b, Gogelein:2010, Gallay:1943}.  Depending on the wetting angle of the secondary fluid, two distinct states are defined. In the pendular state, the secondary fluid preferentially wets the particles and a meniscus forms between particles. The particle-fluid admixture is in the capillary state when the secondary fluid does not preferentially wet the particles. In the pendular state, a sample-spanning network is composed of particles linked by capillary bridges \citep{Koos:2011, McCulfor:2010, Kao:1975b, Cheng:2012}.  In the capillary state, the network consists of particles clustered around small volumes of the secondary fluid \citep{Koos:2012, Fortini:2012}. Capillary suspensions can transition to a gel-like state at volume fractions as low as $\phi$ = 0.10---well below the limit of dense packing---increasing the yield stress and viscosity by several orders of magnitude as the volume fraction of the second fluid is increased \citep{Koos:2011, Cavalier:2002}. These two-fluid suspensions have been created for a variety of material combinations from both water- and oil-continuous suspensions \citep{Koos:2011}.  This two-fluid thickening is even present without the direct addition of a secondary fluid when hygroscopic particles are suspended in oil \citep{Hoffmann:2014}.

Capillary suspensions have a broad spectrum of potential applications \citep{Koos:2011, Koos:2011b}. Capillary suspensions can be used to suspend hydrophilic particles in hydrophobic liquids (and vice versa) without modification of particle surface properties and can prevent the settling of particles \citep{Hoffmann:2014, Eggleton:1954}. Due to the variability in the strength of the capillary force, the phenomenon can be used to adjust the flow properties of coatings, adhesives, or other complex fluids according to the requirements in particular processing steps \citep{Koos:2012b, Narayanan:1989}.
The capillary force usually dominates over other colloidal forces \citep{Seville:2000, Aarons:2005, Leong:1993}, and the strong network found in these suspensions even persists when the major fluid is removed and thus provides a new pathway for the creation of microporous membranes, foams, and ceramics with unprecedented high porosity at small pore sizes \citep{Dittmann:2013}. 
 
Both the capillary state and pendular state are controlled by the capillary force and are strongly influenced by changes in the amount of secondary fluid and the material properties.  In air, the force between two particles depends on the particle radii, the surface tension $\Gamma$ of the fluid spanning the particles, and the wetting angle $\theta$ that fluid makes with the solid surface.  The force between two equally sized spheres of radius $r$ connected by a pendular bridge is given by
\begin{equation}
F_c = 2 \pi r \Gamma \cos\theta 
\end{equation}
where it is assumed that the particles are in contact and that the pendular bridge is small with respect to the radius of the particles \citep{Butt:2009, Herminghaus:2005}. For larger droplets, the force will depend on the volume of the secondary fluid, with various corrections given based on the filling angle and shape of the droplet \citep{Mehrota:1980, Orr:1975, Gao:1997, Willett:2000}. Additional corrections are available for particles that are not in contact or include surface roughness \citep{Lian:1993, Butt:2008, Megias:2009}. 

The behavior of granular materials diverges quickly from a fluid-like behavior with the addition of small amounts of liquid, even from humidity in the air \citep{Herminghaus:2005,Mason:1968, Bocquet:1998, Fraysse:1999}.  These wet granular materials are often described in terms of their liquid saturation, $S$, where a saturation of $S=0$ corresponds to the dry state and $S=1$ to the state where the particles are fully suspended in the liquid.  At low saturations ($S<0.3$), liquid bridges will form between particles with the cohesive force acting through these bridges.  The presence of these bridges will slightly reorganize the mixture resulting in an increase in the porosity \citep{Gogelein:2010}.  With the addition of more liquid, the pores between particles will begin to fill, increasing the cohesion between grains.  The rise in cohesion will continue until almost all of the internal pores are filled but rapidly decreases as full saturation is approached.

For granular materials, the relationship between the capillary force between individual particles and the macroscopic stress for a sample depends on the number of contacts per particle in the agglomerate, the number of particles per unit volume, and the magnitude of the force at each contact.  For a uniform packing of equally sized spheres with liquid bridges between particles in direct contact, the relationship is given by
	\begin{equation}
	\sigma_y = f(\phi) g(\tilde{V_l}) \frac{\Gamma \cos\theta}{r}
	\label{eqn:radius}
	\end{equation}
where $f(\phi)$ is a function of the volume fraction of particles within this sample and $g(\tilde{V_l})$ is a function of the normalized droplet volume $\tilde{V_l}=V_l/r^3$ \citep{Schubert:1984, Pietsch:1967}.  This function $g(\tilde{V_l})$ has a maximum value of unity when the droplets are neither underfilled nor overfilled \citep{Mehrota:1980, Willett:2000}. The particle loading, packing geometry, and the number of particle-particle contacts each influence $f(\phi)$ \citep{Pietsch:1967, Rumpf:1962, Pietsch:1968}.  For low particle volume fractions, as we have in capillary suspensions, $f(\phi) = \phi^2$ would be a reasonable approximation due to the binary nature of particle interactions~\citep{Pietsch:1967}.

In capillary suspensions, the particles are suspended in a bulk fluid rather than in air.  The particles are connected by bridges of the secondary fluid in the pendular state, analogous to wet granular media, but here, the force is proportional to the interfacial tension $\Gamma_{lb}$ and the principal wetting angle is the two-fluid contact angle $\theta_{l,b}$ that the secondary fluid ($l$) makes with the solid particles while surrounded by the bulk fluid ($b$) as shown in Fig. S1 in the supplementary material. For capillary state suspensions, the $\cos \theta$  in eqn.~(\ref{eqn:radius}) has to be replaced by another function of the two-fluid contact angle \citep{Koos:2012}.  The dependence on the interfacial tension and reciprocal radius will remain in place for capillary state suspensions. 

The strength of these capillary suspensions depends on the interfacial tension, contact angle, and drop size.  This drop size depends strongly on the mixing conditions \citep{Domenech:2014} as well as the suspension age and history.  Additionally, the number of particle contacts and particle separation distance may also depend on the suspension age. This current work examines the properties of a capillary state suspension to determine how the network varies with added water content and how this network can age and restructure.  The specific model system used in this work will be discussed in Experiments and methods as well as the experimental methodology.  Experimental results will be included in the Results and discussion section beginning with the results of linear shear experiments followed by the linear viscoelastic properties.  The aging of capillary suspensions will then be discussed at low quasi-static conditions and at higher strain amplitudes. 

\section*{Experiments and methods} \label{exp}
	\subsection*{Materials} \label{exp:mat}

We used hydrophobically modified calcium carbonate suspended in silicone oil ($\phi = 0.11$) with added water in the present work. The calcium carbonate particles were manufactured by Solvay Advanced Functional Minerals (Socal U1S1, Salin de Giraud, France) with an average diameter of 1.6 {\textmu}m as shown in Fig.~S2.  The particle sizes were measured using Laser Diffraction (Sympatec HELOS H0309) when suspended in ethanol and subjected to ultrasonic dispersion (the particles were suspended using a Sympatec QUIXEL unit). The calcium carbonate particles were hydrophobically modified by the manufacturer. 

The samples used throughout this paper were created using thoroughly dried calcium carbonate particles (dried overnight at 70~{\textdegree}C), which were suspended in silicone oil (AK200, Wacker Chemie AG, Burghausen, Germany) with added distilled water. The particles were mixed into the bulk fluid using a turbulent beater blade until a uniform suspension was created (at 500--2,000 rpm for 20 min). This mixture was again dried overnight at 70~{\textdegree}C to remove any water that may have condensed during mixing and held under vacuum (100 mbar) to remove air bubbles. At room temperature, the secondary fluid was added to the suspension and thoroughly mixed using the turbulent beater blade (at 500--1,000 rpm for 10 min). The samples were held in airtight containers until the measurements were completed.  These experimental procedures create homogeneous, repeatable samples and are identical to the preparations used in our previous work on the subject.

Without the addition of any secondary fluid, the calcium carbonate suspension forms a weak van der Waals network in silicone oil with a yield stress of $\sigma_y = 0.5$ Pa. The yield stress is defined as the stress at which the mixture begins to flow, that is, to deform plastically. These measurements were conducted using a stress ramp where the yield stress was found as the point at which slope of the logarithmic deformation (as a function of the logarithmic shear stress) changes from a very low (nonzero) value to a high value. Further confirmation of yield was provided by a drastic reduction in the apparent viscosity. 

The capillary suspensions are characterized using the \emph{saturation} of the preferentially wetting fluid $S$
	\begin{equation}
 	S = \frac{V_\textrm{wetting fluid}}{V_\textrm{total fluid}} = \frac{V_w}{V_b+V_l}
	\end{equation}
which is close to zero for the pendular state ($V_w = V_l$) and approaches one for the capillary state ($V_w = V_b$) \citep{Rumpf:1962, Schubert:1982}.  Here, the symbol $V_l$ represents the total secondary fluid volume in a sample rather than the secondary fluid volume in an individual particle cluster.  The total volume of the bulk fluid is given by $V_b$. 

The mixtures are also characterized by the two-fluid wetting angles, which are calculated from the wetting angles that the individual fluids make with respect to the particles in air.  The fluid wetting angles were determined using a modified Washburn method for the capillary rise of a fluid through a dry powder in air \citep{Gillies:2005} using a DataPhysics DCAT 21 tensiometer (Filderstadt, Germany).  For fluid-particle combinations with contact angles greater than 90{\textdegree} in air, i.e., hydrophobic particles in water, particles were attached to a Wilhelmy plate using double-sided tape and the advancing contact angle was measured \citep{Stiller:2004}.  These values were adjusted to account for errors due to surface roughness following the method of  \cite{Yan:2007} [see also \cite{Wenzel:1936, Nakae:1998}].  Using the values determined in air, the two-liquid wetting angle is calculated using the Young equation \citep{Fournier:2009,Aveyard:2003} as
	\begin{equation}
	\cos\theta_{l,b} = \frac{\Gamma_{la} \cos\theta_{l,a} - \Gamma_{ba} \cos\theta_{b,a}} {\Gamma_{lb}}
	\label{eqn:Young}
	\end{equation}
where the subscript $a$ refers to air, $b$ is the primary (bulk) liquid, and $l$ is the secondary liquid.  The wetting angle $\theta_{\alpha,\beta}$ is the angle fluid $\alpha$ makes against the solid in an environment provided by fluid $\beta$.  The interfacial tension $\Gamma_{\alpha\beta}$ is measured at an interface between the two fluids, as was measured using a Wilhelmy plate on the same tensiometer.  For the material combination used, the contact angle is calculated as $\theta = 123^\circ \pm 5^\circ$.

\subsection*{Measurement procedure} \label{exp:proc}

The dynamic properties of this capillary suspension were studied using stress- and strain-controlled rheometry.  Steady and oscillatory shear measurements were made with a stress-controlled rheometer (Anton-Paar MCR 501) mounted with a plate and Peltier plate geometry of diameter 50~mm (PP50/TG) and gap of 0.5~mm. The plates were covered with a rough grit sandpaper (P40, average roughness of 425~{\textmu}m) to prevent slip (a significant problem for the highly elastic samples). By repeating the measurements at different gap widths, we can confirm that slip does not influence our results. While this rheometer is stress-controlled, the programmed strain waves in a strain-controlled configuration adapted to the correct amplitude within several periods and were purely sinusoidal (as monitored during measurements).  All measurements were made at a temperature $T=20\pm0.01$~{\textdegree}C set via a Peltier plate except where temperature was explicitly varied.  Aging experiments for various samples were also corroborated using an Anton-Paar MCR 702 in 2EC mode (two electrically controlled motors) where the stress was applied using one plate and measured using the other, stationary plate as well as a strain-controlled ARES G2 rheometer (TA instruments) fitted with a stainless steel cone-plate (25 mm, 0.1~rad). 

Our test methodology proceeded as follows. First, a pre-shearing strain rate of $\dot{\gamma} = 200$~s$^{-1}$ was applied for 300~s unless otherwise noted. Such rejuvenation procedures have been shown to fluidize the sample, erasing the history and providing reproducible conditions for the measurement of time-dependent properties~\citep{Ovarlez:2007, Erwin:2010}.   At $t = 0$, the shear stress was set to zero and the sample was allowed to relax. No significant difference in the sample viscosity was observed during consecutive rejuvenation procedures or in the material strength following repeated pre-shearing. Aging measurements were started at $t=0$; other measurements were started after a waiting time $t_w = 1,000$~s.  

Measurements of the sample viscosity were made using a steady shear rate step, where the admixture was sheared at a constant rate until the viscosity reached a steady state.  The time needed to reach such a steady state decreased at higher shear rates, and accordingly, a maximum step time of $t_s = 265 \dot{\gamma}^{-0.37}$~s was chosen. This time was empirically determined using several admixtures and shear rates.  The average of the last ten measurement points was used as the recorded value.

Additional high shear rate measurements were made using a capillary rheometer with piston diameter of 20~mm equipped with a die orifice with a diameter of 0.5- and 15-mm length~\citep{Willenbacher:1997}.  The shear rate is calculated from the adjusted flow volume assuming a Newtonian fluid and neglecting wall slip and pressure loss at the orifice of the die. The shear stress at the die wall is calculated from the measured inner pressure above the die. As the viscosity values for the gelled samples measured with the capillary rheometer agree with the values from rotational rheometry, errors due to slip and entrance effects are assumed to be small and no correction is needed. The values of the 0.00~\% water sample deviate considerably from the data measured with the rotational rheometer. As this deviation increases with lower shear rates where the pressure needed is low, it is assumed that the pressure loss at the die orifice becomes increasingly influential. Therefore, the values of the 0.00~\% sample are Bagley-corrected \citep{Barnes:1989} by repeating the measurement with a die length of 80 mm.

\section*{Results and Discussion}
	
\subsection*{Steady shear} \label{linear}

\begin{figure}[tbp!]
	\includegraphics{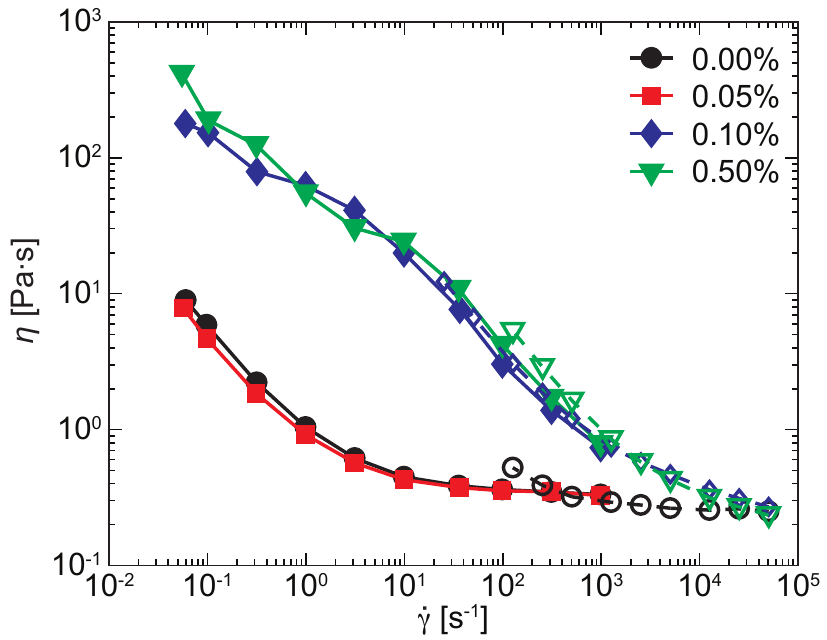}
	\caption{Viscosity of capillary suspension admixtures ($\phi = 0.11$) as a function of shear rate as measured using a rotational rheometer (\emph{filled points with solid lines}) and high shear rate capillary rheometer (\emph{open points with dashed lines})}
	\label{fig:flow_curve}
	\end{figure}

Steady shear viscosity data of the capillary suspension samples were determined in a broad shear rate range covering seven decades.  These measurements, presented in Fig.~\ref{fig:flow_curve}, 
were completed using a rotational rheometer (filled symbols, solid lines) as well as a capillary rheometer (open symbols, dashed lines).  The samples display two disparate regimes based on their water content.  The pure suspension (0.00~\% H$_2$O) and low water content (0.05~\%) admixtures have a much lower viscosity at low shear rates than the higher water content samples (0.10 and 0.50~\%).  All of the samples are shear thinning.  Space-spanning networks are existent in all of the samples due to the van der Waals attraction for the pure suspensions and the stronger capillary network in the high water content samples.  For the low water content samples, the vdW network is partially destroyed at low shear rates ($\dot{\gamma} \approx 10^1$~s$^{-1}$) above which the viscosity remains constant.  For the high water content admixtures, the stronger capillary force can withstand higher shear rates before the agglomerates are completely destroyed and the viscosity drops to the same as the low water content samples.  This high shear rate of $\dot{\gamma}>10^4$~s$^{-1}$ is well beyond the range accessible using a standard rotational rheometer.  This implies that during the rejuvenation segment of any following measurements, the capillary network is not completely eliminated.
	
\begin{figure*}[tbp!]
	\includegraphics{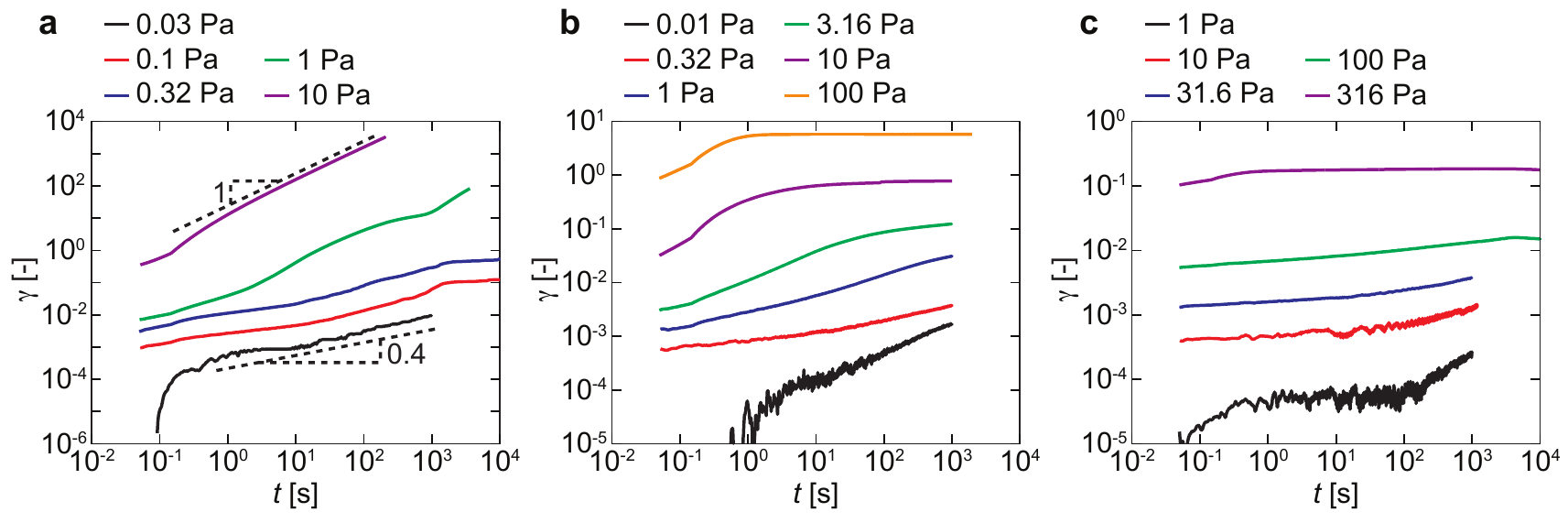}
	\caption{Strain as a function of time for the sample with {\bf a} 0.00, {\bf b} 0.15, and {\bf c} 0.40~\% water for a continuously maintained step-stress. The yield stresses measured for each sample are {\bf a} 0.5 Pa, {\bf b} 110 Pa, and {\bf c} 320 Pa}
	\label{fig:creep}
	\end{figure*}

The differences between samples with low and high amounts of added water are also apparent in creep experiments.  For these experiments, the sample was rejuvenated, kept at rest for 1,000~s, and then subjected to a constant shear stress.  The time $t=0$ represents the beginning of this step-stress.  The response of the sample without added water is shown in Figure~\ref{fig:creep}a.
The sample is freely flowing ($\gamma \propto t$) for applied shear stresses of 10 and 1 Pa (for this sample, the yield stress is $\sigma_y = 0.5$ Pa). For the values below the yield stress, the sample still deforms ($\gamma \propto t^{0.4}$) even at very long times. At very large times, the rate of deformation begins to decrease indicating some resistance of the network to continuing deformation.  At $\sigma=1$~Pa, just above the yield stress of 0.5 Pa, this vdW suspension shows some initial resistance (lower exponent) at very short times before fluid-like behavior is exhibited.  Such a change of the power law exponent is typical of hard sphere and attractive suspensions~\citep{Pham:2008, Erwin:2010}.

The sample with the higher water content (0.40~\% water), shown in Fig.~\ref{fig:creep}c, 
has a yield stress of $\sigma_y = 320$~Pa.  In this case, there is very little deformation exhibited by the sample at all stresses with a slope significantly lower than unity. For this sample, the strain is nearly time invariant even at an applied stress of $\sigma = 316$~Pa, which is just below the yield stress. This very small deformation, even close to the yield stress, indicates the presence of a strong network structure that catastrophically fails when the material yields. Experiments above the yield stress were not performed. 
	
The sample with the intermediate water content (0.15~\% water, $\sigma_y = 110$ Pa), Fig.~\ref{fig:creep}b, 
demonstrates transitional behavior.  The sample flows at short times and for low stresses.  At high stresses, the sample reaches a steady-state deformation.  The time needed to reach this steady state increases with decreasing stress, and below $\sigma = 3$~Pa, no plateau is reached within the experimental time.  This dependence of the time before the samples become arrested is also shown in supplementary Fig. S3 where the measured shear rate is shown as a function of time for each applied shear stress. As with the high water content sample, the network easily resists the highest applied stress of 100~Pa despite the proximity to the yield stress. The difference between the high water content samples and the lower water content samples is more apparent from oscillatory shear measurements. 

\subsection*{Linear viscoelasticity} \label{osc}
\begin{figure}[tbp!]
	\includegraphics{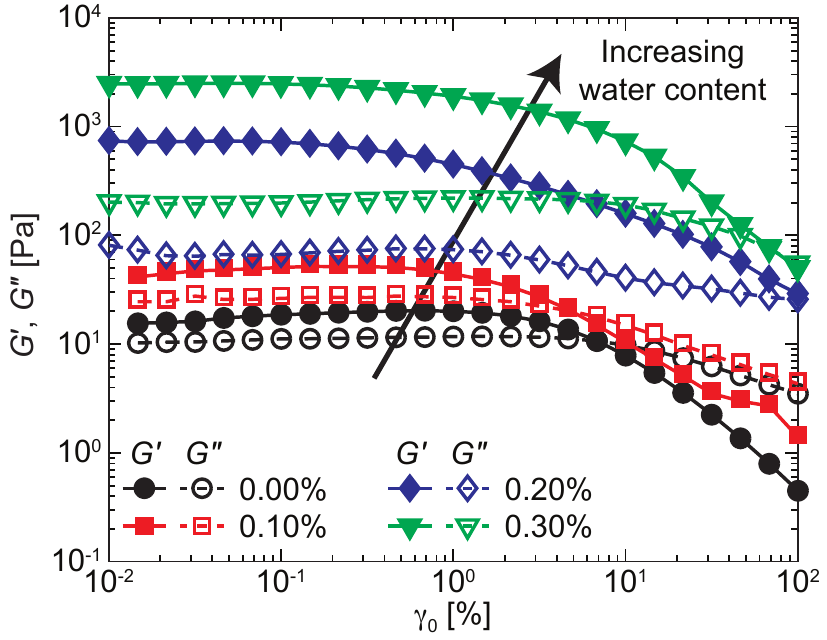}
	\caption{Storage (\emph{filled symbols with solid lines}) and loss (\emph{open symbols with dashed lines}) moduli shown as a function of the deformation amplitude for an oscillation frequency of 1 Hz}
	\label{fig:osc_amp}
	\end{figure}

The capillary suspensions were first characterized using oscillatory shear measurements.  The shear moduli as a function of the oscillation amplitude for a frequency of 1 Hz are shown in Figure~\ref{fig:osc_amp}.
The crossover between $G'$ and $G''$ occurs at lower deformations for the samples with smaller amounts of added water at about $\gamma_y \approx 10~\%$.  For the strong capillary suspensions (e.g., 0.20 and 0.30~\%), the crossover occurs at a deformation of $\gamma_y \approx 100~\%$.  However, the onset of nonlinear response occurs at a much lower deformation $\gamma_c \simeq 0.1~\%$ as judged by the deviation from the low amplitude limiting values.  For the suspension without added water (circles), the moduli remain constant nearly  until the crossover ($\gamma_c \approx \gamma_y$).  In the cluster model for capillary suspensions, where small droplets of secondary fluid are surrounded by tetrahedral or octahedral arrangements of particles which are then networked together as presented in \cite{Koos:2012}, the cluster can sustain some small deformation before the particle separates from the droplet, but the strength of the capillary bridge would depend strongly on this separation distance.  This could give rise to the observed behavior where the crossover is indicative of capillary bridge rupture, but the nonlinearity indicates a dependence of the capillary force on the separation distance.  This nonlinearity can also arise from inhomogeneities in the bridge volumes.  Small bridges will rupture at shorter particle separations~\citep{Willett:2000} decreasing the overall strength of the network. This decrease in strength will continue until a critical number of bridges break and the admixture fluidizes.

\begin{figure}[tbp!]
	\includegraphics{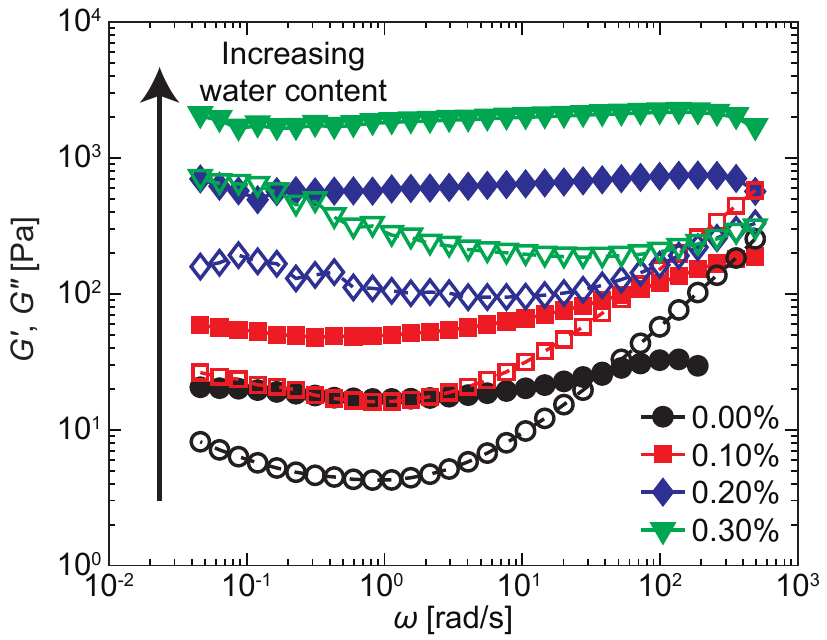}
	\caption{Storage (\emph{filled symbols with solid lines}) and loss (\emph{open symbols with dashed lines}) moduli shown as a function of the oscillation frequency at a deformation of $\gamma_0 =0.1~\%$}
	\label{fig:osc_freq}
	\end{figure}

The transition from the weakly space-spanning network formed by the vdW  to the strong capillary network is readily apparent in Fig.~\ref{fig:osc_freq},
where the shear moduli are shown as a function of the oscillation frequency.  The strength of the moduli is much higher for the capillary suspensions, increasing by three orders of magnitude with addition of 0.40~\% added water to the suspension of calcium carbonate in silicone oil.  This dramatic increase in the strength of the material is typical for capillary suspensions.  For the weakly aggregated suspensions, there is a crossover such that the viscous behavior dominates at high frequencies.  This crossover shifts toward higher frequencies with increasing water contents. For the capillary suspensions, $G'$ remains much higher than $G''$ for all frequencies tested here, but there is presumably still a crossover.  If the trend displayed in the last few points continues, this crossover will occur at approximately 2,000~rad~s$^{-1}$ and $1\times10^4$~rad~s$^{-1}$ for the 0.20 and 0.30~\% samples, respectively.  This crossover must be confirmed with higher frequency measurements. 

\subsection*{Aging at rest} \label{aging}

\begin{figure}[tbp!]
	\includegraphics{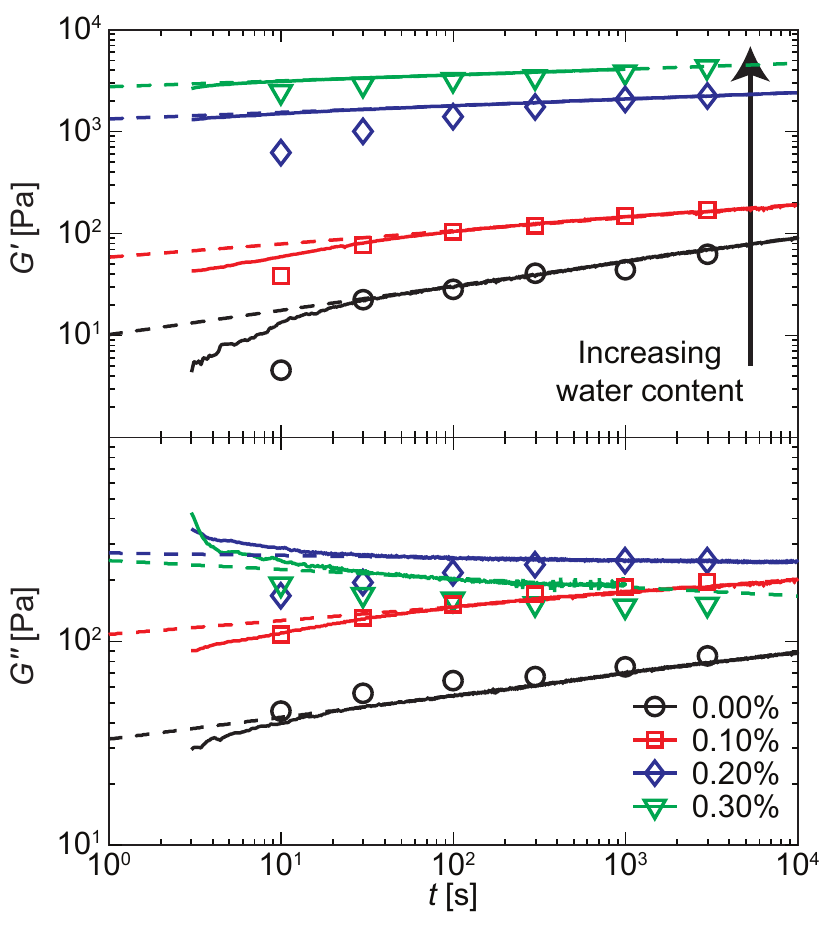}
	\caption{Storage and loss moduli as a function of the time since network rejuvenation. Measurements subjected to constant oscillatory deformation $\gamma_0 = 0.1~\%$ at 10 Hz) are shown as \emph{solid lines}. Points are for measurements that are only deformed at the specified times and are at rest for the remaining time. \emph{Dashed lines} represent power law fits}
	\label{fig:normal_recovery}
	\end{figure}

The strength of capillary suspensions has also been observed to vary in time~\citep{Koos:2012} indicating network rearranging.  In this previous work, a difference between quiescent aging and aging monitored with continuous oscillation was observed.  As shown in Fig.~\ref{fig:osc_amp}, the end of the linear viscoelastic regime occurs at very low oscillation amplitudes.  To test if the previously observed aging behavior was due to the higher amplitude used, this test is repeated at a much lower amplitude as shown in Fig.~\ref{fig:normal_recovery}.
After rejuvenation, the admixtures are allowed to age quiescently with periodic monitoring or with continuous monitoring.  The shear modulus is measured using a constant deformation amplitude of $\gamma_0 = 0.1~\%$ at 10 Hz.  The results for both quiescent and constant oscillatory aging agree implying that that this very small deformation does not affect structural reorganization. The storage modulus $G'$ increases with time for all of the samples but the modulus increases more rapidly for the lower water content samples than for the higher water content samples.  For the loss modulus $G''$, the strength increases with time for the lower water content samples, but decreases for the higher water content samples.  Such a reduction in $G''$ is associated with a reduction of disorder within the sample following gelation of suspensions \citep{Ovarlez:2007}.  This transition from a loss modulus that increases in time to one that decreases occurs for added water between 0.15 and 0.20~\% furthering the idea that these samples represent an intermediate state between the van der Waals dominated behavior and capillary suspensions.

\begin{figure}[tbp!]
	\includegraphics{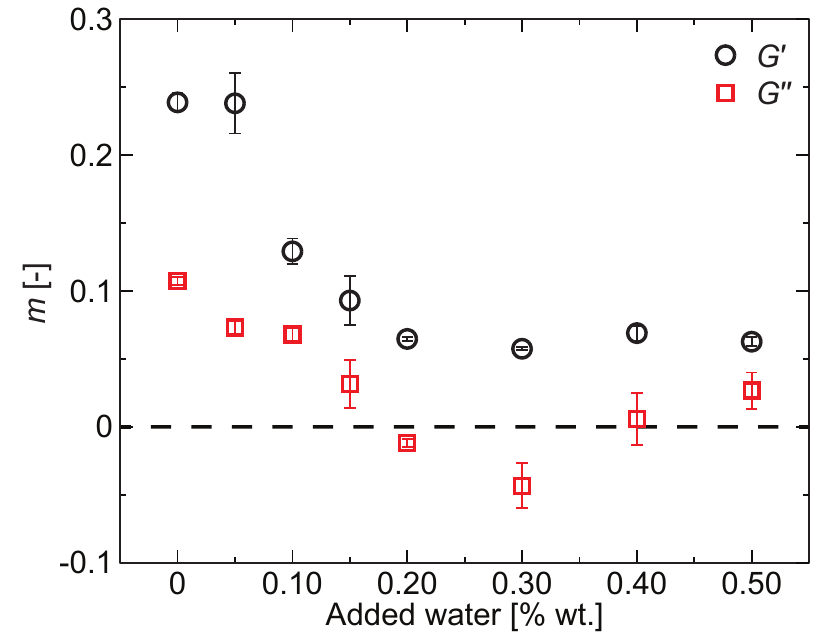}
	\caption{Exponents from power law fits obtained for storage and loss moduli data as a function of the added secondary fluid content.  \emph{Error bars} represent the standard errors for each fit}
	\label{fig:pl_params}
	\end{figure}

The storage and loss moduli both appear to follow a power-law aging process continuing for very long times as has been shown for other soft materials~\citep{Fielding:2000}.  As such, the data was fit with a power law of the form
	\begin{equation}
	G = \left(\frac{t}{\tau} \right)^m
	\label{eqn:pl}
	\end{equation}
using a nonlinear least squares method for $t \geq 10$~s.  These fits are shown in Fig.~\ref{fig:normal_recovery} as dashed lines, and the fitting parameter $m$ is shown in Fig.~\ref{fig:pl_params}.
The exponent $m$ decreases rapidly as small percentages of water are added to the sample transforming the mixture from a vdW-dominated to capillary-dominated behavior. For the high water content samples, the exponent for the storage modulus remains fairly constant meaning that the aging behavior is uniform for each of the four samples but may be shifted up or down based on the initial material strength.  The exponent for the loss modulus also decreases from the higher value for samples without added water, obtains a minimum value for a water content of 0.30~\%, and then increases at the highest water contents. For the suspension without added water, the aging is best fit with $G' =  \left( t/1.74 \textrm{ s} \right)^{0.24}$ and $G'' =  \left( t/1.45 \textrm{ s} \right)^{0.11}$ whereas the admixture with 0.30~\% is best fit with $G' =  \left( t/1.58 \textrm{ s} \right)^{0.06}$ and $G'' =  \left( t/0.79 \textrm{ s} \right)^{-0.04}$.  For the intermediate water contents where the $G''$ exponent is negative, this may imply that free particles become attached to the network or that the microstructural configuration is modified to reinforce some connections. These modifications will increase the network strength, increasing $G'$ while decreasing $G''$, during the experimental time. The change in the aging behavior of $G''$ while $G'$ remains fairly constant may imply that the network structure becomes more branched for increasing water contents where these branches do not contribute to the overall network strength. If there exists a more branched network structure in the high water content sample, the binding of a free particle to a side branch would become increasingly likely.

 While our samples appear to age according to a power law, we know that this fit cannot hold for very short or very long times.  Instead, the admixture should age slowly, then rapidly, and then slowly again as it approaches its long-time steady-state value as the system undergoes structural arrest and further vitrifies. The monotonic process typically observed can either be a consequence of a limited observation window or inadequate rejuvenation (as is likely the case in the current system).  A sigmoidal model following the form of \cite{Negi:2010} was also tested, but as the very short time network formation response and the plateau at very long times are not captured, this fit yields no new information about the material.  The sample without added water does reach a plateau after 6~h ($2.16 \times 10^4$~s).  In contrast, the 0.50~\% water content sample follows the power law aging of eqn.~(\ref{eqn:pl}) for over 60~h ($2.16 \times 10^5$~s). In this case, the experiment failed due to the onset of slip triggered by external vibrations before a plateau was reached.

\subsection*{Shear-mediated aging} \label{aging:amp}

\begin{figure}[tbp!]
	\includegraphics{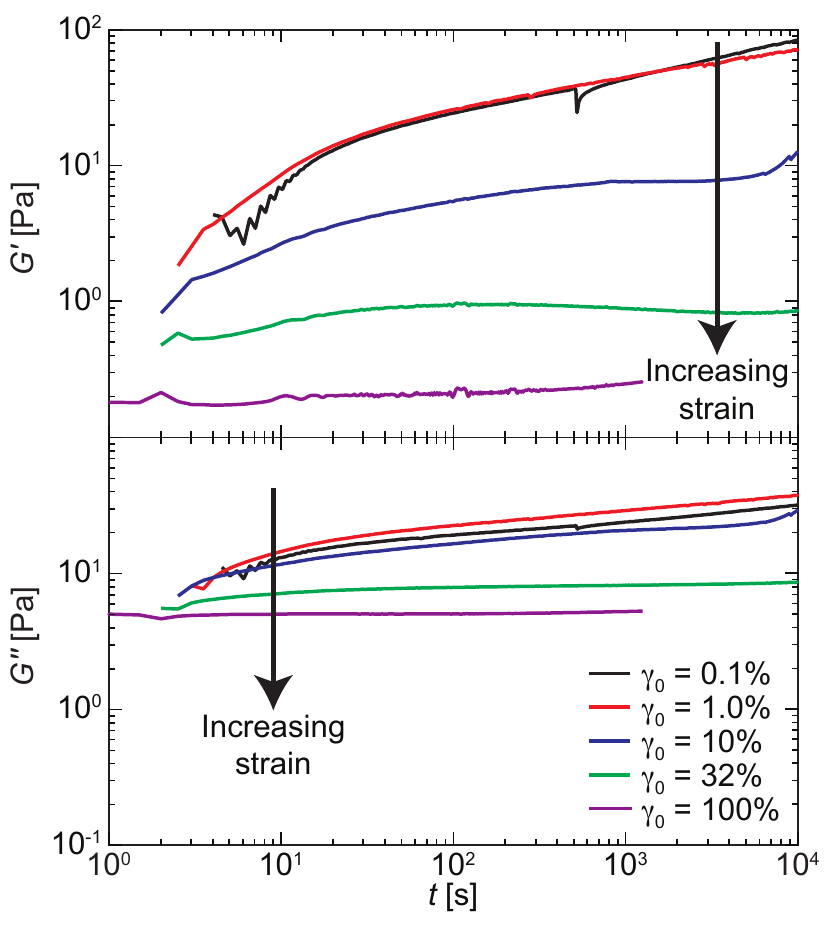}
	\caption{Storage and loss moduli as a function of the time since network rejuvenation for the sample with 0.00~\% added water.  Measurements were subjected to a constant oscillatory deformation between 0.1 and 100~\% at a frequency of 1~Hz}
	\label{fig:amp00}
	\end{figure}

The strength of the capillary network shows a time-dependent response as the sample is allowed to age on the rheometer following a rejuvenation period and also shows a strong dependence on the oscillation amplitude. Thus, it is hypothesized that the aging of these materials can vary based on the applied deformation amplitude.  For the sample without added water, the strength of both the storage and loss moduli increases in time for low $\gamma_0$, but the slope decreases with increasing oscillation amplitude and eventually $G'$ and $G''$ remain constant for $\gamma_0=100~\%$ as shown in Fig.~\ref{fig:amp00}. 
There is also a slight decrease in the storage modulus observed for the lowest oscillation amplitude at around 1,000~s. This slight change is not caused by errors due to external vibration but is present in many samples and may indicate some network reformation. This must be tested however through observations of the network structure.

\begin{figure}[tbp!]
	\includegraphics{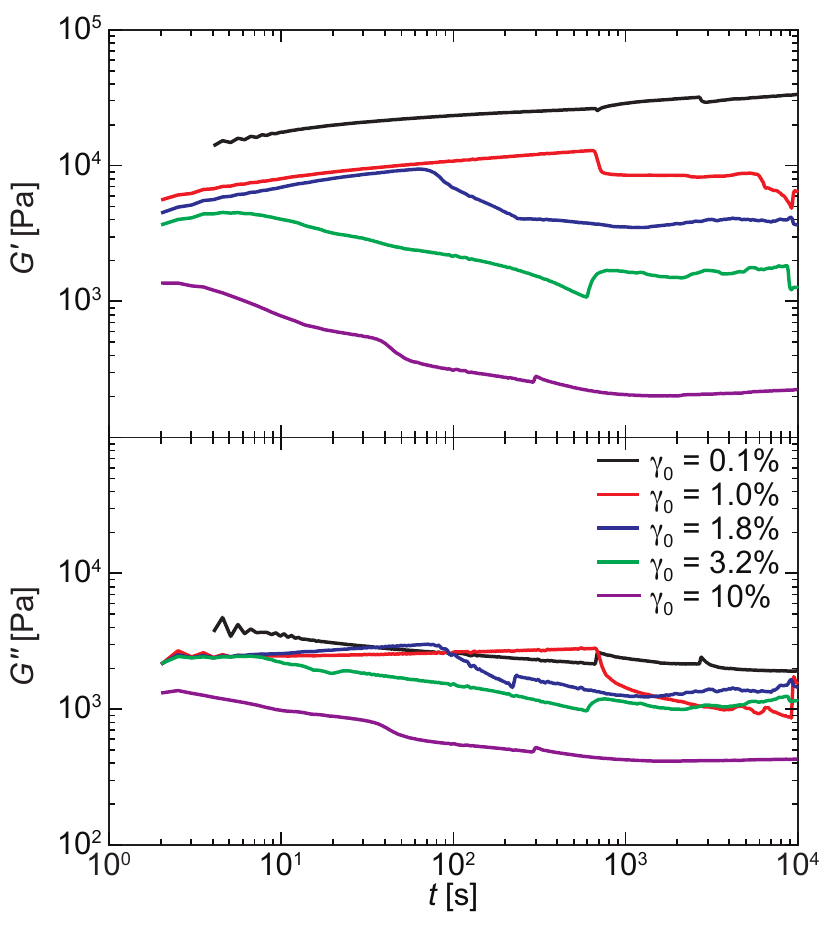}
	\caption{Storage and loss moduli as a function of the time since network rejuvenation for the sample with 0.30~\% added water.  Measurements were subjected to a constant oscillatory deformation between 0.1 and 10~\% at a frequency of 1~Hz}
	\label{fig:amp30}
	\end{figure}

Capillary suspensions exhibit a different behavior, as demonstrated for the 0.30~\% sample (Fig.~\ref{fig:amp30}).
At low deformations, the storage modulus increases and the loss modulus decreases in time.  As the deformation amplitude is increased slightly, the storage modulus begins to increase with a similar slope as for the lowest deformation, but then at some critical time, it suddenly decreases and never fully recovers.  The loss modulus transitions from continually decreasing to remaining constant (e.g., $\gamma_0 = 1.0~\%$) or even increasing slightly (e.g. $\gamma_0 = 1.8~\%$).  At the point where the storage modulus decreases, the loss modulus also decreases.  This transition point occurs at approximately the same time for multiple repetitions and multiple sample batches and indicates some rupture of the capillary network. Both moduli continuously decrease at high deformation amplitudes.  While these amplitudes are below the crossover amplitude $\gamma_y$, they are well above the linear viscoelastic region $\gamma_c$.  

\begin{figure}[tbp!]
	\includegraphics{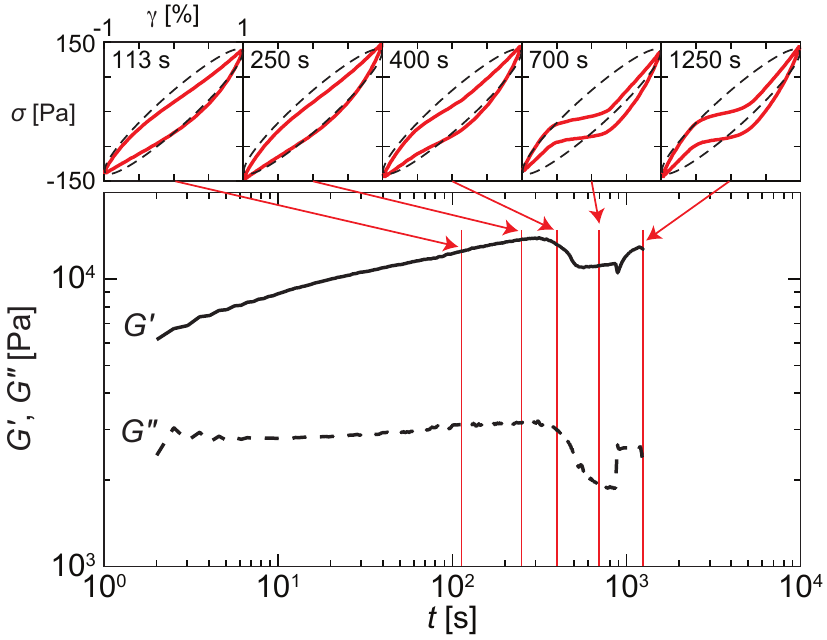}
	\caption{Storage (\emph{solid line}) and loss moduli (\emph{dashed line}) as a function of the time since network rejuvenation for the sample with 0.30~\% added water ($\gamma_0 = 1~\%$, $f = 1$ Hz) (\emph{bottom}). Lissajous plots for at times indicated with elliptical paths shown as \emph{dashed lines} (\emph{top})}
	\label{fig:lissajous}
	\end{figure}

The material response is examined near the drop in moduli to ascertain what is happening in the material near this critical point as shown in Fig.~\ref{fig:lissajous}.
Well before the drop, the Lissajous plot does not have an elliptical path, but the nonlinearity becomes increasingly large as time progresses (for comparison, the Lissajous data at increasing strain amplitudes is shown in the supplementary material). This nonelliptical behavior increases significantly after the drop in $G'$ and remains high even at the last time shown, where the moduli appear to recover slightly. If the sudden drop in $G'$ is associated with fracturing of the capillary network, this persistent nonlinear behavior makes sense, as the network cannot recover while it is subjected to this high deformation.  Prior to fracture, the capillary bridges will be stretched and may rearrange into a slightly reinforced state. In the model for capillary state suspensions presented in \cite{Koos:2012}, the network is composed of particles organized into a tetrahedron around a small drop of secondary fluid.  These tetrahedrons can be joined along a facet (three shared particles) or vertex (one shared particle).  During the shear-mediated aging, the configuration can change so that  the network is joined by the tetrahedral facets rather than vertices.

In capillary suspensions, the capillary force supplied by the secondary fluid droplets controls the attractive interactions between adjacent particles.  There will be a distribution in droplet sizes corresponding to a range in interparticle forces.  While this distribution can be made narrower through high-speed mixing \mbox{\citep{Domenech:2014}}, there will still be weak interactions in the network that are prone to failure.  These weak connections are typical of wet granular media \mbox{\citep{Rahbari:2009}}, which shows some key similarities with capillary suspensions.  As an applied force is propagated through the network, the network is able to stretch or rearrange, with each individual particle deformation depending on the volume of the connecting bridges and on the neighboring interactions. While this deformation should result in the majority of particle contacts becoming stronger (corresponding to the increase in network strength observed), it can also weaken some bridges.  These weak bridges will deform more than the adjacent strong contacts (perhaps leading to the increasing nonlinearity observed in Fig.~\ref{fig:lissajous}) until they rupture.  This rupture is exacerbated by the inhomogeneities in droplet sizes so that there is a distribution in interparticle forces.

\begin{figure}[tbp!]
	\includegraphics{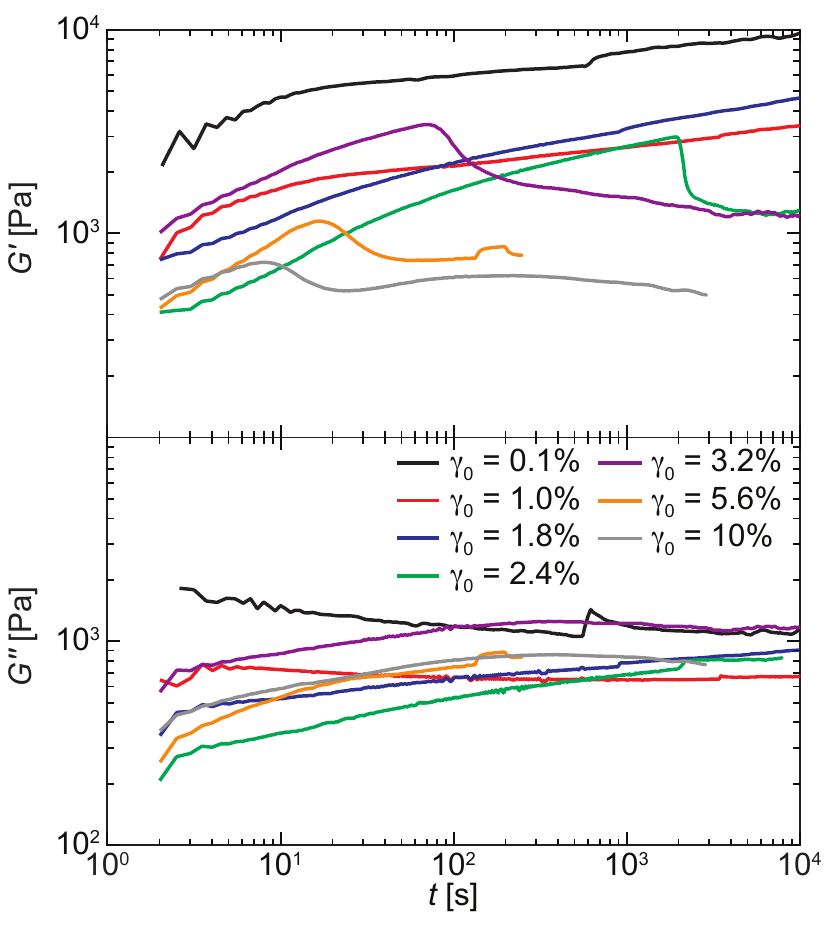}
	\caption{Storage and loss moduli as a function of the time since network rejuvenation for the sample with 0.15~\% added water.  Measurements were subjected to a constant oscillatory deformation between 0.1 and 10~\% at a frequency of 1 Hz}
	\label{fig:amp15}
	\end{figure}

For the sample with 0.15~\% added water (Fig.~\ref{fig:amp15}),
the low amplitude behavior ($\gamma_0 = 0.1~\%$ and 1.0~\%) resembles the vdW suspensions for $G'$ where the modulus constantly increases and capillary suspensions for $G''$ where the modulus continuously decreases. This is reversed for deformations above the linear viscoelastic regime ($\gamma_0 \geq 1.8~\%$) where the storage modulus behaves as observed for the capillary suspensions: It begins to increase before a sudden drop is observed. The loss modulus for these high deformations behaves like the low water content vdW suspensions in that the modulus continually increases.  At the time associated with sudden drop in $G'$, no change in $G''$ is observed.

\section*{Conclusions} \label{concl}
This paper describes how capillary suspensions respond to external forces using a calcium carbonate suspension with added water.  This suspension is a capillary suspension in the capillary state~\citep{Koos:2011}.  Without added water, a weak space-spanning network is formed by the attractive van der Waals force.  With an added water content of 0.20~\%, a strong network dominated by the capillary force is created. Intermediate behavior is observed for water contents less than 0.20~\%.  

The vdW network and the associated clusters are very sensitive to external deformations demonstrating a nearly continuous deformation even at stresses well below the yield stress. The capillary suspension easily resists constant external shear stresses, rapidly obtaining a constant deformation even at stresses very close to the yield stress.  Clusters formed from this strong capillary force persist at very high shear rates ($\dot{\gamma}<10^4$~s$^{-1}$) in contrast to the vdW clusters that are destroyed at shear rates $\dot{\gamma} \geq 10^1$~s$^{-1}$. Intermediate water contents form a network composed of the strong clusters with the associated strong resistance to external shear rates, but a network that deforms for short times and low stresses before a constant deformation value is reached.

Under oscillatory shear deformation, the end of the linear viscoelastic region corresponds to the crossover deformation ($\gamma_c \approx \gamma_y \approx 10~\%$). The capillary suspensions have a much lower critical deformation ($\gamma_c \approx 0.1~\%$) but higher crossover ($\gamma_y \approx 100~\%$). The force of attraction in capillary suspensions is strongly dependent on the separation between particles and the size of the bridges.  This dependence is clearly demonstrated for the difference in $\gamma_c$ and $\gamma_y$ for the samples with high water contents. This sensitivity to applied deformation may be due strength of the capillary bridges decreasing in strength with the particle separation before rupturing. Alternatively, inhomogeneities in droplet sizes may allow small bridges to rupture at smaller deformations decreasing the shear modulus prior to the fluidization at the crossover deformation. These differences between capillary suspensions and suspension without added water are also demonstrated with oscillation frequency where the weakly aggregated suspensions demonstrate a crossover at lower frequencies than for the strong capillary suspensions.

The strength of this network can be increased through the sample age after rejuvenation and has been monitored using very small oscillatory deformations.   Both capillary suspensions and the van der Waals suspensions can restructure and age for very long times and are well fit by a power law.  The power law exponent is higher for the vdW suspensions and continually decreases with increasing water contents until it reaches a constant value at 0.20~\%. In capillary suspensions, larger deformations cause a modification of the shear moduli and suggest at network rearrangement and rupture. This rearrangement and rupture is magnified by inhomogeneities in the droplet volumes with mixed vdW and capillary forces such that rupture occurs when the weakest force-chain is broken. These changes in the sample-spanning network will have to be confirmed with direct observations.

\begin{acknowledgements}
EK would like to acknowledge financial support from the European Research Council under the European Union's Seventh Framework Program (FP/2007-2013)/ERC Grant Agreement no. 335380.
\end{acknowledgements}

\bibliographystyle{RA}      

\end{document}